# Client applications and Server Side docker for management of RNASeq and/or VariantSeq workflows and pipelines of the GPRO Suite

Hafez, A.[1,2,3#], Soriano, B.[1,4,5#], Elsayed, A.A.[1,6#], Futami, R.[1], Ceprian, R.[1], Ramos-Ruiz, R.[7], Martinez, G.[1], Roig, FJ.[8], Torres-Font, M.A.[1,9], Naya-Català, F.[4], Calduch-Giner, J.A.[4], Trilla-Fuertes, L.[10], Gamez-Pozo, A.[10,11], Arnau, V.[5], Sempere, J.M.[6], Pérez-Sánchez, J.[4], Gabaldón, T.[2,12,13,14], and Llorens, C.[1*]

# These authors contributed equally

1.- Biotechvana, Parc Científic Universitat de València

2.- Universitat Pompeu Fabra, Barcelona, Spain

3.- Faculty of Computers and Information, Minia University, Egypt

4.- Institute of Aquaculture Torre de la Sal (IATS), CSIC, Castellón, Spain

5.- Institute for Integrative Systems Biology (I2SysBio), UVEG-CSIC. Valencia, Spain.

6.- Valencian Research Institute for Artificial Intelligence (VRAIN-UPV), Universitat Politècnica de València, Valencia, Spain

7.- Genomics Unit Cantoblanco, Parque Científico de Madrid, Madrid, Spain

8.- Facultad de Ciencias de la Salud, Universidad San Jorge, Zaragoza, Spain

9.- Escola Tècnica Superior d'Enginyeria, Universitat de València, Spain

10.- Biomedica Molecular Medicine SL, Madrid, Spain

11.- Molecular Oncology and Pathology Lab, Hospital Universitario La Paz-IdiPAZ, Madrid, Spain

12.- Institució Catalana de Recerca i Estudis Avançats, Barcelona, Spain

13.- Barcelona Supercomputing Centre (BSC-CNS). Barcelona, Spain

14.- Institute for Research in Biomedicine (IRB), The Barcelona Institute of Science and Technology, Barcelona, Spain

***Corresponding authors:**

carlos.llorens@biotechvana.com

**Abstract**

The GPRO suite is an in-progress bioinformatic project for -omic data analyses. As part of the continued growth of this project, we introduce a client side & server side solution for comparative transcriptomics and analysis of variants. The client side consists of two Java applications called *“RNASeq”* and *“VariantSeq”* to manage pipelines and workflows for RNA-seq and Variant-seq analysis, respectively, based on the most common command line interface tools for each topic. As such, *“RNASeq” and “VariantSeq”* are coupled with a Linux server infrastructure (named GPRO Server Side) that hosts all dependencies of each application (scripts, databases, and command line interface software). Implementation of the server side requires a Linux operating system, PHP, SQL, Python, bash scripting, and third-party software. The GPRO Server Side can be deployed via a Docker container that can be installed in the user’s PC using any operating system or on remote servers as a cloud solution. *“RNASeq”* and *“VariantSeq”* are both available as desktop (RCP compilation) and cloud applications (RAP compilation). Each application has two execution modes: a Step-by-Step mode enables each step of the workflow to be executed independently and a Pipeline mode allows all steps to be run sequentially. *“RNASeq”* and *“VariantSeq”* also feature an experimental, online support system called GENIE that consists of a virtual (chatbot) assistant and a pipeline jobs panel coupled with an expert system. The chatbot can troubleshoot issues with the usage of each tool, the pipeline job panel provides information about the status of each computational job executed in the GPRO Server Side, and the expert provides the user with a potential recommendation to identify or fix failed analyses. *“RNASeq”* and *“VariantSeq”* and the GPRO Server Side combine the user-friendliness and security of client software with the efficiency of front-end & back-end solutions to manage command line interface software for RNA-seq and variant-seq analysis via interface environments.

## Introduction

Advances in next generation sequencing (NGS) have changed the way researchers perform comparative analyses based on RNA-seq and variant-seq resequencing data (for a review, see for example [1]). Nevertheless, implementing these approaches into routine laboratory procedures remains challenging as they require the sequential execution of complex and variable protocols to extract and process the biologically relevant information from the raw sequence data. These protocols are typically called pipelines and/or workflows and are usually carried out using command line interface (CLI) software. The advantage of these pipelines is that they can be customized for specific goals and utilize the wide range of freely available CLI software produced by the scientific community. This is particularly useful for resequencing RNA-seq and variant-seq approaches where the requirements of each pipeline will differ depending on the data to be analyzed. For example, RNA-seq pipelines vary depending on the availability of GTF/GFF files (the file format that provides information about the gene features of a reference sequence) and the reference sequence (it can be a genome, a transcriptome, a gene panel, etc.). Similarly, variant-seq pipelines vary depending on the type of variants (single point mutations, indels, etc.) or according to the source and frequency of the target variants (somatic or germinal). Another advantage of protocols based on CLI tools is that they run on both personal computers (PCs) and computational servers. This allows the simultaneous management and analysis of multiple samples, a practice that is typical in RNA-seq and variant-seq approaches. The disadvantages of pipelines based on CLI tools is that their implementation and usage can only be achieve on Linux environments and requires advanced informatic skills for installing third-party software, writing scripts, and executing processes with the command line. In other words, these protocols are restricted to experienced bioinformaticians.

In recent years, many Graphical User Interface (GUI) have been developed to provide user-friendly tools for NGS data analysis. Most those focusing on RNA-seq and variant-seq are cross-platform desktop applications distributed under payment licenses (for a review see [2]). These applications are typically implemented under intuitive and secure frameworks but, in comparison to pipelines based on CLI tools, they are significantly limited in terms of analytical tasks and are less efficient than pipelines for processing multiple samples (as it is typical in RNA-seq or variant-seq studies). The advantage of desktop applications is thus their ease-of-use, which only requires informatic skills at the user level. However, they are not as efficient or versatile as pipeline or workflow protocols assembled from CLI tools. As such, an effective strategy is to provide end-users with GUIs for managing CLI tools via web servers that apply front-end & back-end programming. Examples of front-end & back-end solutions are the bioinformatic databases and online repositories like Ensembl [3], the NCBI web resources [4] or projects like Galaxy [5] a platform of precompiled web modules adapted to manage CLI tools in GUI environments. Galaxy modules can be combined to construct and personalize workflows and pipelines for RNA-seq, variant-seq, and/or for any other -omic approach (the repertoire of tools supported by Galaxy project is indeed extensive and impressive). However, implementing a Galaxy solution remains complex and installing and configuring specific combinations of Galaxy modules requires advanced bioinformatic skills with a significant background in informatic systems.

With the aim to address the previously mentioned issues, we launched GPRO a bioinformatic project whose initial release (1.n) was a multi-task desktop application [6,7] with client functions to perform functional analyses via cloud computing strategies. The second and current version (2.n) consists of a suite of applications each devoted to a specific topic. In this article, we describe a new client side & server side solution for this suite to perform comparative transcriptomics and variant analysis. We have published a describe for another application for the GPRO suite, *"SeqEditor"* – an application for sequence analysis[8], and the remaining application will be published in future articles. More information about the GPRO project is available at https://gpro.biotechvana.com. The here introduced solution consists of two client applications named *"RNASeq"* and *"VariantSeq"* and a bioinformatic server platform called GPRO Server Side (GSS) that is coupled to the client applications. The GSS contains the CLI tools, databases, and other dependencies needed by *"RNASeq"* and *"VariantSeq"* to assemble pipeline and workflow protocols. The GPRO suit also features a smart experimental artificial intelligence system for user support called GENIE, which will also introduced in this article.

## Material and Methods

### Client Side applications

The framework of *“RNASeq”* and *“VariantSeq”* was developed in Java and Desktop and Cloud versions were created using Eclipse Rich Client Platform (RCP) and Eclipse Remote Application Platform (RAP) respectively [9]. The implementation of this framework follows a similar approach to Model–view–controller (MVC) pattern [10]. At the model layer, the framework includes all implementations needed to represent low level element of the tools’ wrapper descriptor (e.g. JobDescriptor and different types of VariableDescriptor such as input files and tools parameters) as well as workflow templates’ descriptors. At the view layer, we implemented automated utilities to generate GUIs for single tasks or workflows within each CLI tool using the selected JobDescriptor or WorkflowTemplates. At the controller level, the implementation includes the Task or workflow instances controlling and storing user inputs captured by the GUIs based on the model layer and that are also responsible for executing and tracking the tasks on the GSS. As part of the workflow framework at the controller layer, the Bash framework validates the tasks from the user side and generates bash scripts from tasks descriptors submitting them to the GSS for running. In such scripts, tracking events are inserted to track general tasks, check the status of running tasks, and collect log files. All events are stored in user space on the GSS and are sent back to the client’s applications for visualization.

**GPRO Server Side Platform.**

GSS is a Linux infrastructure that hosts all the dependencies required by *"RNASeq"* and "*VariantSeq"* to run pipelines and workflows on the server side. GSS constitutes of the following elements:

- Linux Operating System with at least Bash version 4
- MySQL Server for installing databases
- Apache HTTP server 2.2 or later
- PHP 7 or later
- R and Bioconductor
- Perl 5
- Python 2.7
- Third party CLI software (see Table 1 for details)
- An API for communicating between client applications and GSS.

Installation of GSS requires complex steps to setup Linux, Apache, MySQL, and PHP (LAMP stack) as well as the CLI software. It also requires scripts for handling the incoming requests to GSS that must be manually installed. To overcome this, we have deployed GSS in a Docker container [11] that can be easily installed on remote servers or any PC or Mac using the OS, Windows, Linux as long as there is sufficient disk space and RAM. Minimum requirements are 500 Gb of hard disk and 16Gb of RAM.

Table 1.- CLI software dependencies of "*RNASeq*" and *"VariantSeq"* at the GSS

| | **CLI third party software** | ***RNASeq*** | ***VariantSeq*** |
|---|---|---|---|
| Quality Analysis and Preprocessing | FastQC [v0.11.5] [12] | ✓ | ✓ |
| | FastqMidCleaner [1.0.0] | ✓ | ✓ |
| | Cutadapt [1.18] [13] | ✓ | ✓ |
| | Prinseq [PRINSEQ-lite 0.20.4] [14] | ✓ | ✓ |
| | Trimmomatic [0.36] [15] | ✓ | ✓ |
| | FastxToolkit [0.0.13] [16] | ✓ | ✓ |
| | FastqCollapser [1.0.0] | ✓ | ✓ |
| | FastqIntersect (1.0.0) | ✓ | ✓ |
| Mapping of reference genome or transcriptome | TopHat [v2.1.1] [17] | ✓ | ✓ |
| | Hisat2 [2.2.1] [18] | ✓ | ✓ |
| | Bowtie2 [2.2.9] [19] | ✓ | ✓ |
| | BWA [0.7.15-r1140] [20] | ✓ | ✓ |
| | STAR [2.7.0f] [21] | X | ✓ |
| Quantification | Corset [1.06] [22] | ✓ | X |
| | Htseq [0.12.4] [23] | ✓ | X |
| Post Processing | Bed Tools [v2.29.2] [24] | X | ✓ |
| | GATK [v4.1.2.0] [25,26] | X | ✓ |
| | Picard tools [2.19.0] [27] | X | ✓ |
| | SAMtools [1.8] [28] | X | ✓ |
| Transcriptome Assembly | Cufflinks [v2.2.1] [29] | ✓ | X |
| Differential Expression | DESeq [2.1.28] [30] | ✓ | X |
| | EdgeR [3.30.3] [31] | ✓ | X |
| | Cuffdiff [v2.2.1] [29] | ✓ | X |
| | CummeRbund [2.30.0] [32] | ✓ | X |
| Enrichment Analysis | GOseq [1.40.0] [33] | ✓ | X |
| Training Sets | GATK [25,26] | X | ✓ |
| Variant Calling | GATK) [v4.1.2.0] [25,26,34] | X | ✓ |
| | VarScan2 [v2.4.3] [35] | X | ✓ |
| Variant Filtering | GATK [v4.1.2.0] [25,26] | X | ✓ |
| Annotation of variant effects | Variant Effect Predictor [105.0] [36] | X | ✓ |

"✓" means yes and "X" means not included. All the CLI software here summarized integrated in the GSS docker image excepting Varscan2 because licensing questions. Authors interested in VarScan2 may find indications about how also install this tool in the GSS docker at https://gpro.biotechvana.com/tool/gpro-server/manual. Academic users can freely install it while commercial users need to contact the VarScan2 authors to get the corresponding commercial license (for more details see https://github.com/dkoboldt/varscan/releases).

**Virtual chatbot assistant and expert system**

*"RNASeq"* and *"VariantSeq"* are supported by an experimental, artificial intelligence (AI) system called GENIE that was created and trained using natural language processing and machine learning methodologies [37,38]. GENIE consists of distinct interfaces, dialogs, and scripts (the client side part) that are linked to a server side module composed of the following elements: 1) knowledge databases; 2) the expert system; and 3) the virtual (chatbot) assistant. These three features are centralized in a GPRO remote server so that the expert system and chatbot can be continually fed new training data. Below is a detailed description of each element.

Knowledge databases: The chatbot and the expert system are supported by five knowledge databases that are shared between the virtual assistant and the expert system:

- Questions & answers database. This database identifies and stores key terms and serves as an index of answers to different questions.
- CLI tools dependency database. This database stores information on the type of input that each CLI tool receives and the output that it generates, as well as information of different parameters and customization options.
- Contextual database. This database provides a graphical representation to all pipelines/workflows and the programs implemented in each protocol.
- Key terms database. This is a database of generic questions about different protocols or programs.
- Log files database. This is a database that stores the information reported

by the log files generated by the CLI software dependencies.

Information for these databases was taken from the *“RNASeq”* and *“VariantSeq”* manuals (available at the Section “Data Availability Statement”) and from public scientific networks and/or repositories such as Biostar [39], SeqAnswers [40], Pubmed [41] and the GATK community forum (https://gatk.broadinstitute.org/hc/en-us/community/topics).

Expert system: This is a rule-based system that provides users with actionable solutions for troubleshooting problems in failed analyses. The expert system was implemented in Python using the Django framework (https://www.djangoproject.com) and trained using machine-learning methodologies [42,43]. It consists of::

- Inference engine: This handles the users’ request by processing the logs and the tracking information sent by the job tracking panel of client applications with the objective of extracting key features and errors information that can be used to query the solutions database.
- Proven facts database: This database contains the rules managed by the inference engine for recommendations of how to fix problems and errors from failed analyses.
- Administration panel: This is a web site provided for administration and management of the expert system when applying rules or adjusting aspects such as adding new task descriptors, editing databases, managing actions/recommendation templates, etc. The administration panel is only accessible by experts from our side or by users interested in contributing to the training of this tool.

- Client interface: This is the interface implemented in the pipeline jobs panel of the client applications (“*RNASeq”* and *“VariantSeq”*) to manage the interaction with the expert system engine.
- API: The API allows the interface to accept requests from the client applications and enables client applications to track and fetch the actions/recommendations proposed by the expert system.

Chabot Engine: The chatbot helps users to resolve issues with installation, technical errors, user guides, or FAQs. The chatbot engine was implemented using python via the Rasa open-source framework [37] and pre-trained Universal Sentence Encoder language models [38]. The chatbot engine utilizes a Retrieval-based strategy with intent classification, entity identification and extraction, and response selection from a set of predefined responses. The chatbot is considered as a level 3 conversational AI as it can understand questions from the context and handle unexpected queries (users changing their mind, etc.). The training dataset was mainly compiled from our collection of Q/A database focusing on client applications and bioinformatic related concepts and extended to other Q/A data sources (the about referred knowledge databases). Users are allowed to interact with the chatbot via two different interfaces:

- Online Web interface available at https://gpro.biotechvana.com/genie. This webpage includes a dialog where users can ask questions and the chatbot will respond using a graphical summarization of the different protocols of each GPRO application including "*RNASeq*" and "*VariantSeq*".

- An interactive user interface implemented in each client application to query the chatbot directly from the application.

The chatbot allows an API developed using the Rasa framework to modulate the communication between client applications and the chatbot.

## Results

### General overview

"*RNASeq*" and "*VariantSeq*" are two cross-platform client applications built for the processing and analysis of resequencing data obtained via NGS technologies. Specifically, "*RNASeq*" offers a GUI-based environment to manage pipelines and workflows based on CLI tools for differential expression (DE) and enrichment analysis. "*VariantSeq*" offers a similar solution but for calling and annotation of single point mutations (SNP) and indels. "*RNASeq*" and "*VariantSeq*" can be installed on the user's PC (desktop version) or used via web browser (cloud or web version). Analyses performed by "*RNASeq*" and "*VariantSeq*" are executed in GSS via a Linux server infrastructure hosting a collection of CLI tools (Table 1) used by both applications as pipelines and workflows dependencies. To this extent, GSS includes an API and other server side dependencies needed to link each client application ("*RNASeq*" or "*VariantSeq*") to GSS. Figure 1 shows a technical schematic for the framework of "*RNASeq*" or "*VariantSeq*" and how it operates for executing single analyses or pipeline complex analyses in GSS. As the latter is a complex infrastructure, it has been deployed in a docker container that can be easily installed on remote servers or the user's PC. The current version of the GSS docker supports one or two users working simultaneously; however, we are committed to releasing a future version for servers with multiple users. Currently, servers with requirements for multiple users will have to install GSS manually tool-by-tool (server administrators interested in that possibility can contact us for more detailed information).

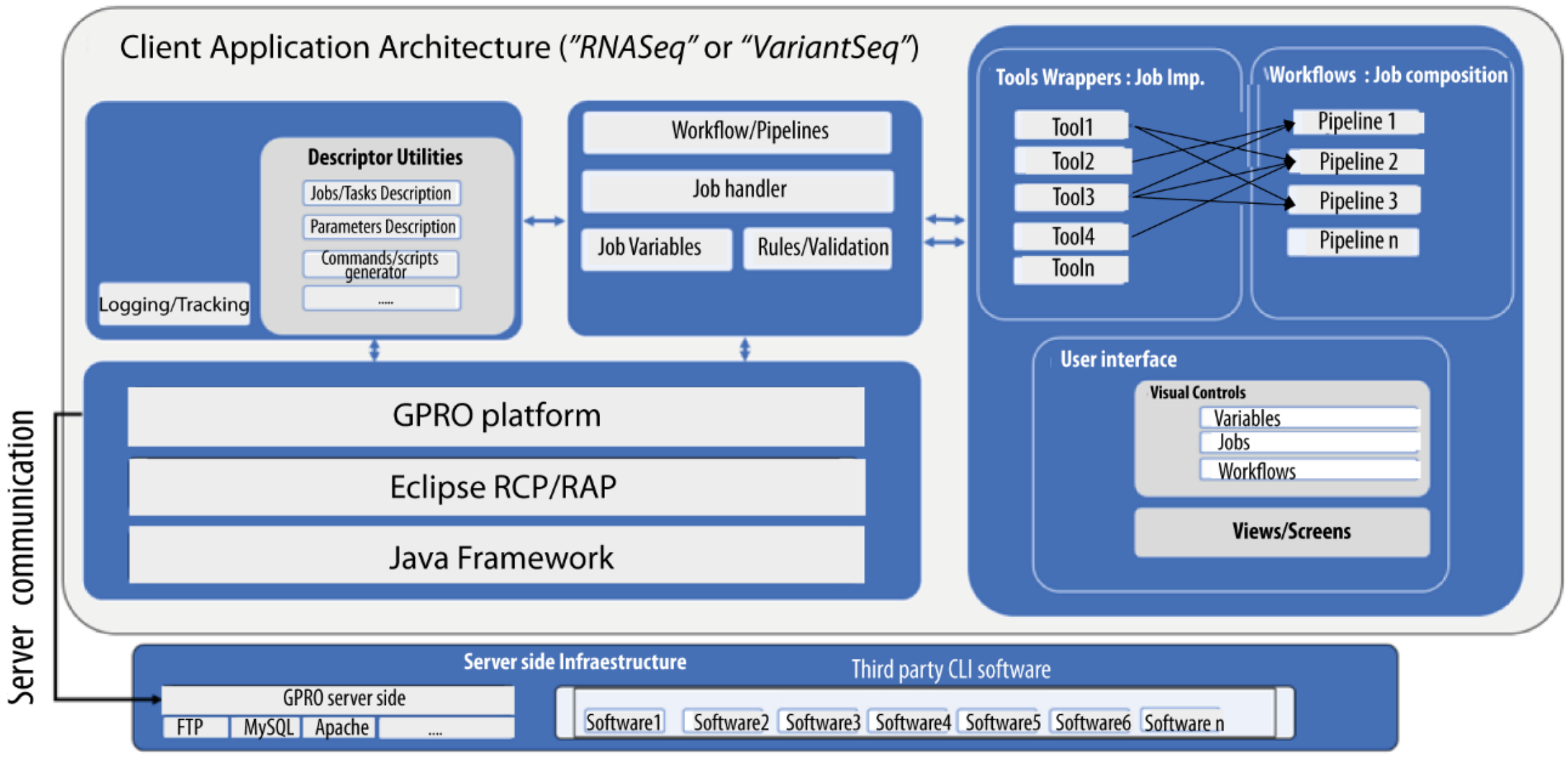

**Figure 1.- Client side and server side schematic and implementation.** *"RNASeq"* and *"VariantSeq"* were both implemented using a common eclipse framework that enables encapsulation of third-party CLI tools as task wrappers, dynamically generated GUI views for each CLI tool, executable scripts, composable pipelines, and tracking/logging outputs of running jobs. The GSS provides the Linux environment and all other server requirements to run the CLI software (including scripts, R, Perl, Python, and MySQL server). Applications and the GSS connect via API.

## User interface

*"RNASeq"* and *"VariantSeq"* use a common user interface (shown in Figure 2) to access the GSS and manage analyses. The user interface is structured into the following modules:

- *“FTP Browser.”* This is a File Transfer Protocol (FTP) to provide users access to the GSS and to transfer files/folders from the user PC to the GSS, or vice versa.
- *“Working space.”* This is the framework space from which the GUIs manage the CLI tools hosted at the GSS.
- *“Top Menu.”* This is the main menu for each application and is located at the top of the interface. All tools and tasks are organized into different tabs as detailed below:
    - *“Directory.”* This tab is for users to select and set the main directory for exchanging material with the GSS using the FTP browser.
    - *“Transcripts/Variant Protocols.”* This tab provides access to the modes of computation and protocols of each application. By clicking on this tab, the user can choose between two computational modes: Step-by-Step or Pipeline. When selecting the Step-by-Step mode, a *“Task Menu”* appears in the working space to provide access to the set of GUIs for the distinct CLI tools and/or commands implemented in the step-by-step workflow for each application. When choosing the Pipeline mode, the user accesses to the pipeline manager of each application.
    - “*Pipeline Jobs.*” This tab allows the user to track the status of all jobs executed in the GSS or to obtain recommendations from the GENIE’s expert system to troubleshoot computational issues in failed analyses.
    - *“Preferences.”* This tab allows the user to configure and activate the connection settings between the client application and the GSS.

- *“Help.”* This tab provides access to the user manual for each application and to the summary panel of the GENIE’s chatbot.

As previously noted, analyses are run on the server side and so the client applications and GSS must be linked. To do this, users must access *“Pipeline connection settings”* in the “Preferences” tab and configure the connection settings as illustrated in Supplementary file S1.

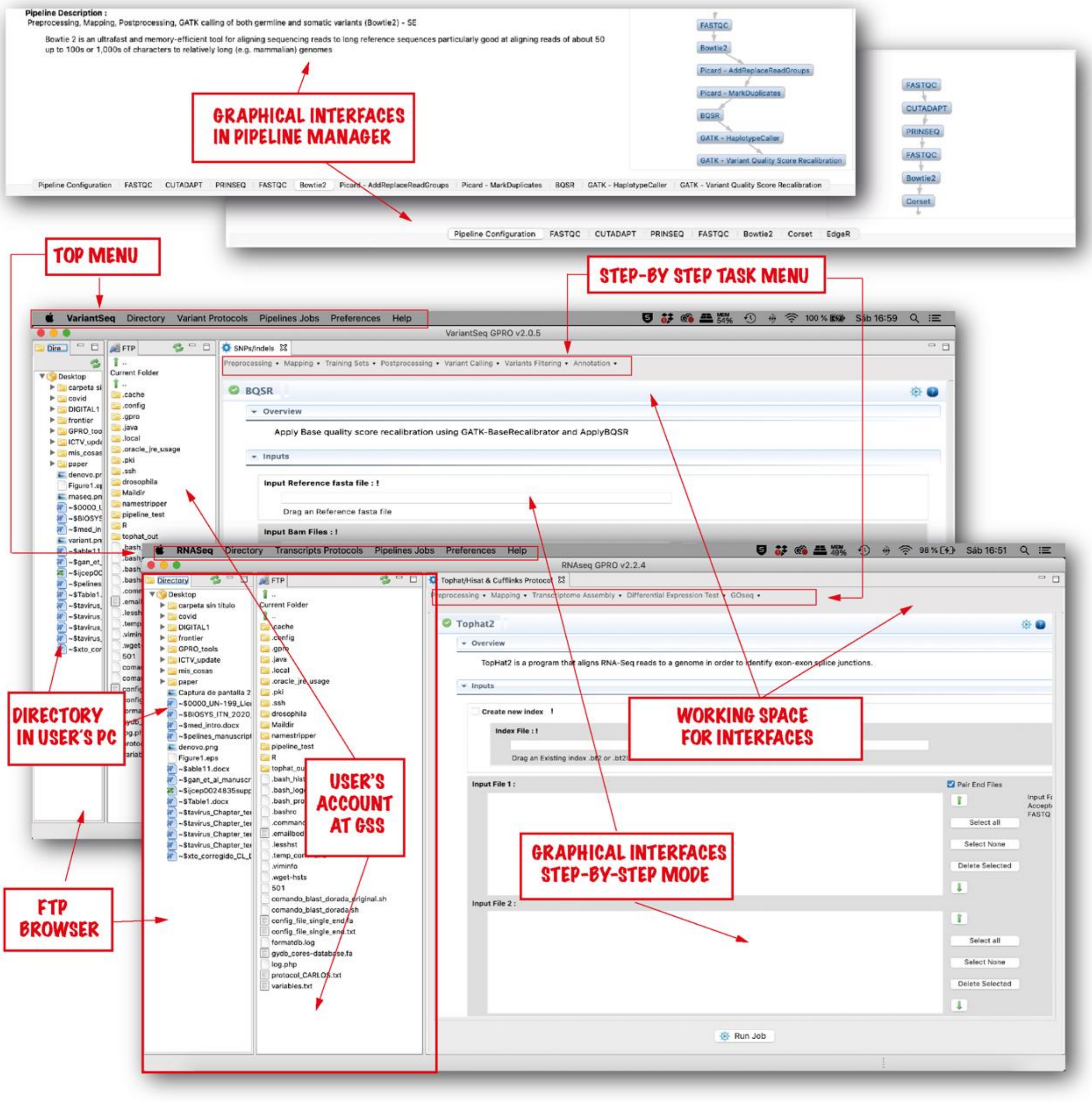

**Figure2**.- **User interfaces of *“RNASeq”* and *“VariantSeq”.*** Both applications have a common interface. The general procedure is as follows. Once an application has been linked to the GSS, the user should follow the subsequent steps. (1) Transfer the input files from user´s PC directory to the GSS using the FTP browser. (2) Select the computational mode (it can be step-by-step or pipeline-like) (3) Drag the input files (fastq files, reference sequences, GTF/GFF files, Training Sets, etc) from the GSS to the input fields of the selected interface/s. (4) declare the output. (5) Set options and parameters. (6) Run the analysis.

**Protocols**

*"RNASeq" and "VariantSeq"* have been created based on two "good-practice" protocols for the most common and popular CLI tools in each topic (for more details see the following reviews [44-48]). In Figure 3, we show the protocol for DE and enrichment analysis based on which *"RNASeq"* has been implemented. This protocol is based on the following steps: "*Quality Analysis & Preprocessing"* where distinct tools for quality analysis and preprocessing of fastq samples are provided, *"Mapping"* offering tools to map the reads of fastq files against reference sequences, *"Transcriptome Assembly* and/or *Quantification"* to assemble and quantify the transcriptome expression patterns of the case study samples by processing the bam files obtained at the mapping step, *"Differential Expression"* for comparison for the distinct groups/conditions under comparison, *"Differential Enrichment"* for assessing differential enrichment of Gene Ontology (GO) categories and/or metabolic pathways. Two possible paths are allowed within this protocol. One path follows the "*Tophat-Cufflink"* good-practices [29] where splicing mappers such as Tophat or Hisat2 [18] are combined with the Cufflinks package [29,32] to perform splicing mapping and DE analyses. These are mainly oriented (but no limited) to RNA-seq studies using genome sequence references usually accompanied with GTF/GFF files. The other path is a *"Mapping & Counting"* protocol, where DNA/RNA mappers such as Bowtie [19], BWA [20] or STAR [21] are combined with tools for transcriptome quantification like Corset [22] or HtSeq [23] to perform DE analysis with EdgeR [31] and DESeq2 [30]. This path is usually used in RNA-seq studies based on sequence references with no availability of GTF/GFF files such as transcriptomes assembled de novo, amplicons, and gene sets. Under both paths, we consider a final of differential enrichment of GO categories and/or metabolic pathways using GOSeq [33].

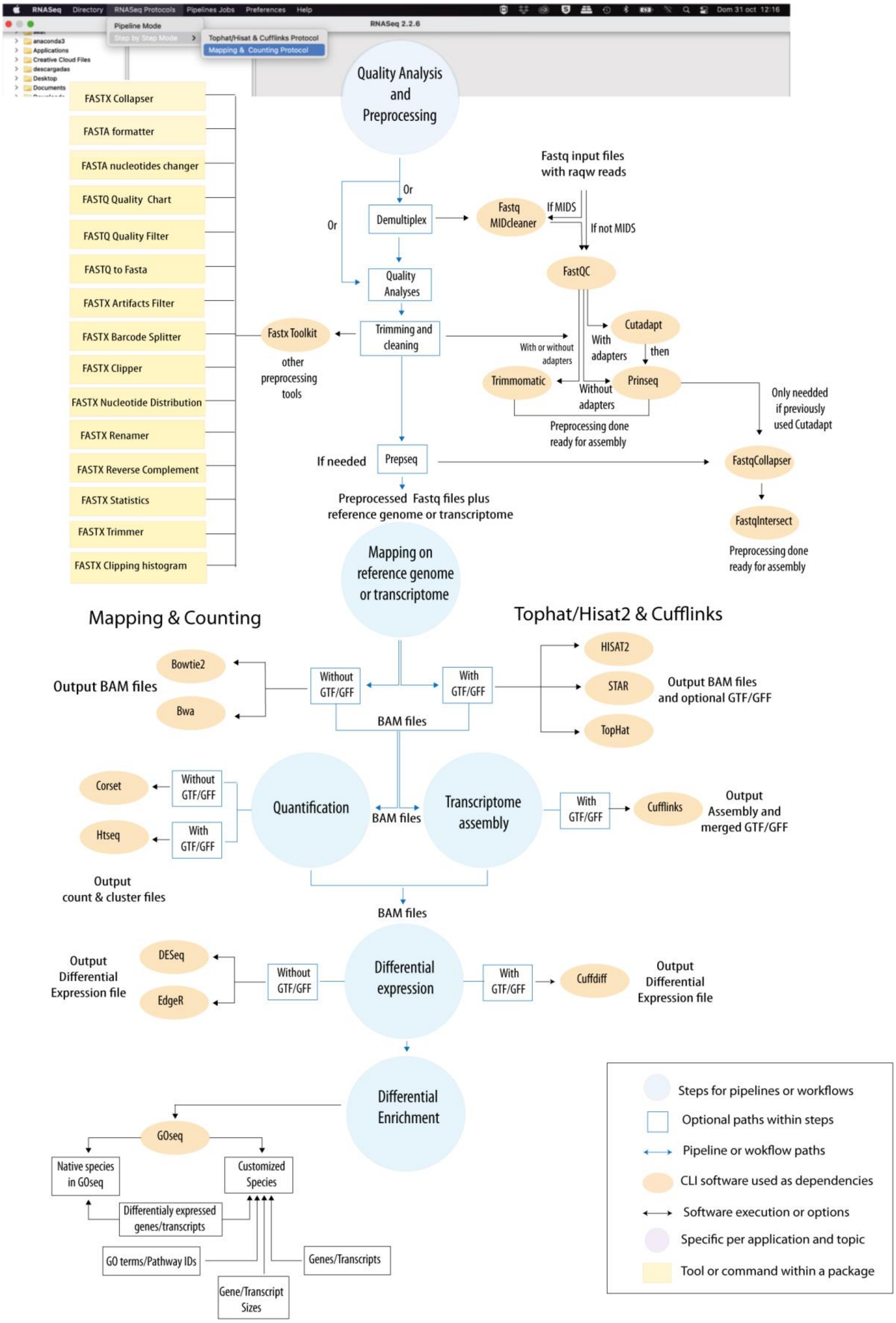

**Figure 3.- *"RNASeq"* protocol.** Computational steps constituting the protocol of *"RNASeq"* for DE and enrichment analysis. The protocol is based on the following steps: *"Quality Analysis & Preprocessing", "Mapping"; "Quantification";*

*“Transcriptome Assembly”; “Differential Expression” and “Differential Enrichment.”* A summary of All CLI tools available for each step is provided in the figure. Two alternative paths (respectively designated as “Mapping & Counting” and “Tophat/Hisat2 & Cufflinks”) are allowed.

*“VariantSeq”* was developed following a protocol based on the most common practices for calling/annotation of SNP and indels using the GATK [25,26] and VarScan2 [35] callers and other CLI tools, including Picard [27], SAMtools [28] and others. As shown in Figure 4, the protocol of “VariantSeq" presents the following steps; *“Quality Analysis & Preprocessing”* for fastq files preprocessing*; “Mapping”* for mapping fastq files against reference sequences; *“Training Sets”* for generating computational resources such as panels of normals (PON), training sets, truth sets, and known-site sets usually required to eliminate or reduce false positives; *“Postprocessing”* for preparing bam files for the calling step (by marking duplicates, re-aligning reads/variants, adding tags, indexing data, etc); *Variant Calling”* for performing the variant calling using different command options of the GATK and VarScan2 packages that will vary depending on the data in analysis (genome, exome, transcriptome) and type of variant (germinal, somatic, cancer, trio, etc); *“Variant Filtering”* for postprocessing VCF files generated in the variant calling steps to filter the variants according to different criteria like coverage, quality, frequency; *“Annotation”* for processing VCF files to add functional effect annotations to the called variants using the Variant Effect Predictor (VEP) of Ensembl [36].

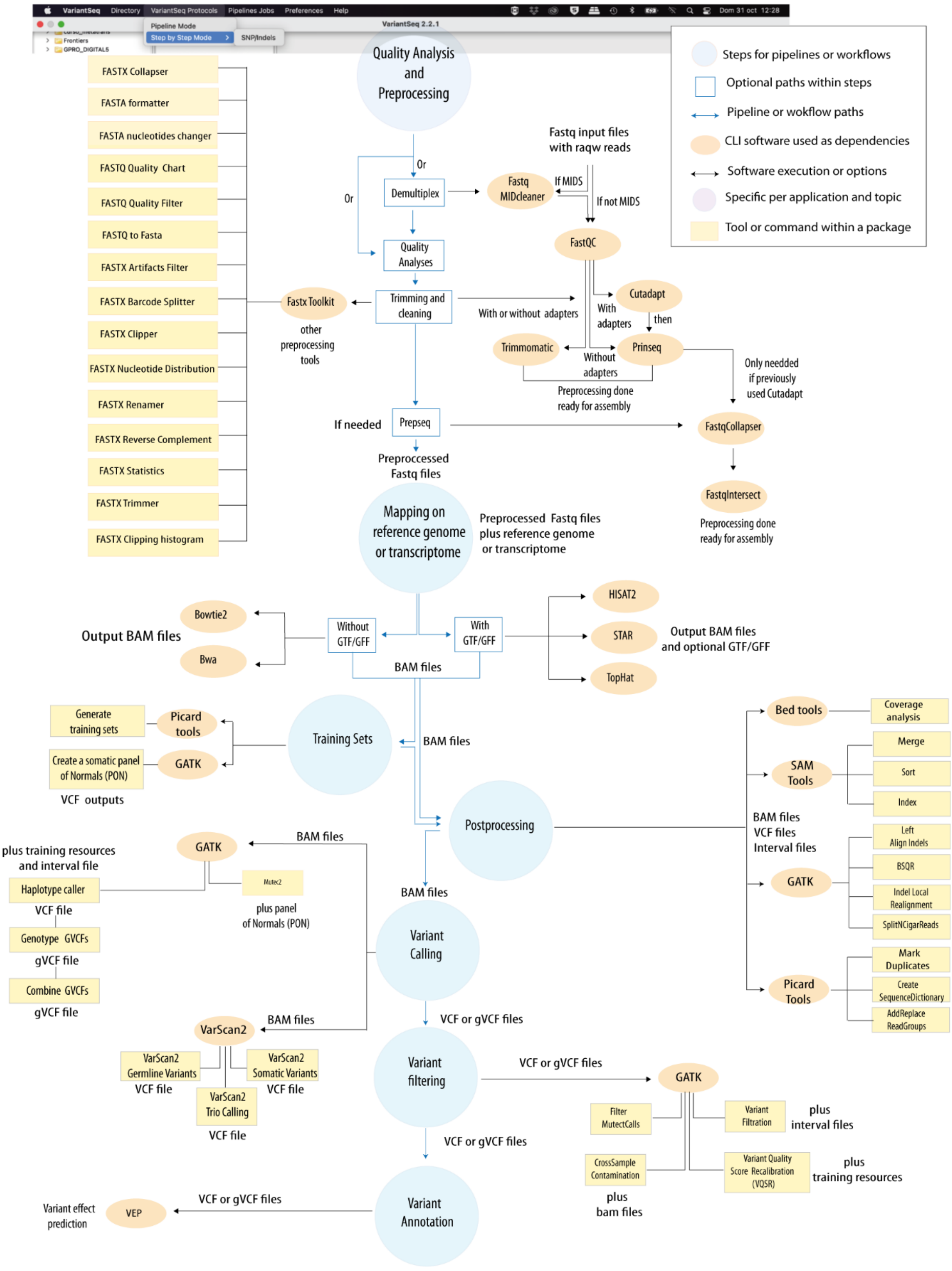

**Figure 4.- "*VariantSeq*" protocol.** Computational steps within *"VariantSeq"* for calling and annotation of SNP and indels. This protocol is based on these steps; *"Quality Analysis & Preprocessing"; "Mapping"; "Training Sets"; "Postprocessing"; "Variant Calling"; "Variant Filtering";* and *"Annotation"*. Depending on the combination of tools selected in this protocol users can call distinct types of variants (germinal, somatic, cancer, trio) from genome, exome, transcriptome NGS data.

**Usage and tutorials**

*"RNASeq"* and *"VariantSeq"* can be executed using two different modes: Step-by-Step or Pipeline. In *Step-by-Step* mode, analyses are executed in a stepwise manner and each analysis can be executed and/or re-executed independently from all other analyses. The *"Task Menu"* will appear in the working space with several tabs organized according to the steps of the workflow. Each tab has a scroll down sub-menu with distinct options and CLI tools to perform any analyses associated with that step. In Step-by-step mode, each CLI tool has a GUI for users to declare input and output files, configure options and parameters, and run the analyses as could be achieved using the command line. Two examples of GUIs per CLI tool are provided in Supplementary Files S2 and S3 (one from "*RNASeq"* and another from *"VariantSeq"*). In the Pipeline mode, users can access the Pipeline manager of each application to configure and run specific, sequential combinations of CLI tools. When the user accesses the Pipeline manager a summary with all possible pipeline combinations appears allowing users to select one of these pipelines. Next, the user access another interface to upload the input data files and output folders and for setting the experiment design (identifying the groups/conditions to be compared or declaring which fastq files are replicates of a group or condition, etc). Afer this, users can access a pipeline menu where they can configure the options and parameters of each CLI tool associated to each step of the pipeline. Once the pipeline is configured, the user can run all the steps of the analyses in one click. In Supplementary Files S4 and S5, we present two dynamic gifs that Illustrating the procedure to configure and run the respective pipeline managers of *"RNASeq"* and *"VariantSeq"*. In addition, two tutorials are available on the installation and usage of *"RNASeq"* and *"VariantSeq"* using real data from previously published works. One is a *"RNASeq"* tutorial based on a control vs infection case study of comparative transcriptomics performed by Pérez-Sanchez et al. [49] on sea bream *Sparus aurata*. The other tutorial is for *"VariantSeq,"* and it is based on a case study of cancer variant analysis previously performed by Trilla-Fuertes et al. [50] using whole-exome data sequenced from human anal squamous cell carcinoma. These two tutorials are freely available in the web sites of the manuals of *"RNASeq"* and *"VariantSeq"*. A direct link to each tutorial is also provided in the section below "*Data Availability Statement".*

**User support system**

*"RNASeq"*, *"VariantSeq",* and GSS are linked to a smart system called GENIE that provides each application with two support tools: i) a pipeline jobs panel powered by an expert system to monitor the status of all pipeline jobs submitted to the GSS and for providing users with recommendations to fix failed analyses; ii) a virtual chatbot assistant to answer questions related to the usage of each application, protocols, and features of each CLI tool. In Figure 5, we provide a technical schematic of the GENIE system and screenshots of the chatbot and the Pipeline Jobs panel. The knowledge databases and engine cores of the chatbot and the expert system are hosted on a remote server of the GPRO project. This allows for the centralized training, growing, curation and continual improvement of these AI systems. Each application implements dialogs and panels that interact with GENIE via API. The interface dialog for interacting with chatbot is accessible in the "*Help"* section of each application albeit a web version of this dialog is also available online at https://gpro.biotechvana.com/genie. The pipeline jobs panel is a dynamic register that allows the user to monitor and review the history of each job submitted to the GSS. As shown in Figure 5, this panel is structured into three screens: i) A top screen showing all job/pipeline records submitted to the GSS; ii) A middle screen showing all track information for a selected job record; iii) A bottom screen showing the log file (stdout and stderr) of the executed job. The history shown in the pipeline jobs panel is periodically updated and users can also update this manually via the context menu. By right clicking on any history record, users have access to a contextual menu allowing the following tasks:

- *"Select in FTP Explorer".* This opens/views the output folder of the selected record.
- *"View Report".* This visualizes the log file of the selected record.
- *"Refresh".* This manually refreshes the history records.
- *"Delete".* This deletes the selected record from the history (this only deletes the record and cached log and track information. The original files with the results are kept on the server and can only be deleted directly from the server or from the FTP Browser).
- *"Restart".* This runs the analysis again with the same input data options and parameters used in the previous analysis.
- *"Edit & Restart".* This runs the analysis again but allows the user to edit or modify any input data, option, or parameter from the previously used CLI tool.
- *"Resolve".* This accesses the interface of the expert system allowing to provide recommendations on controlled actions as defined by the expert system.

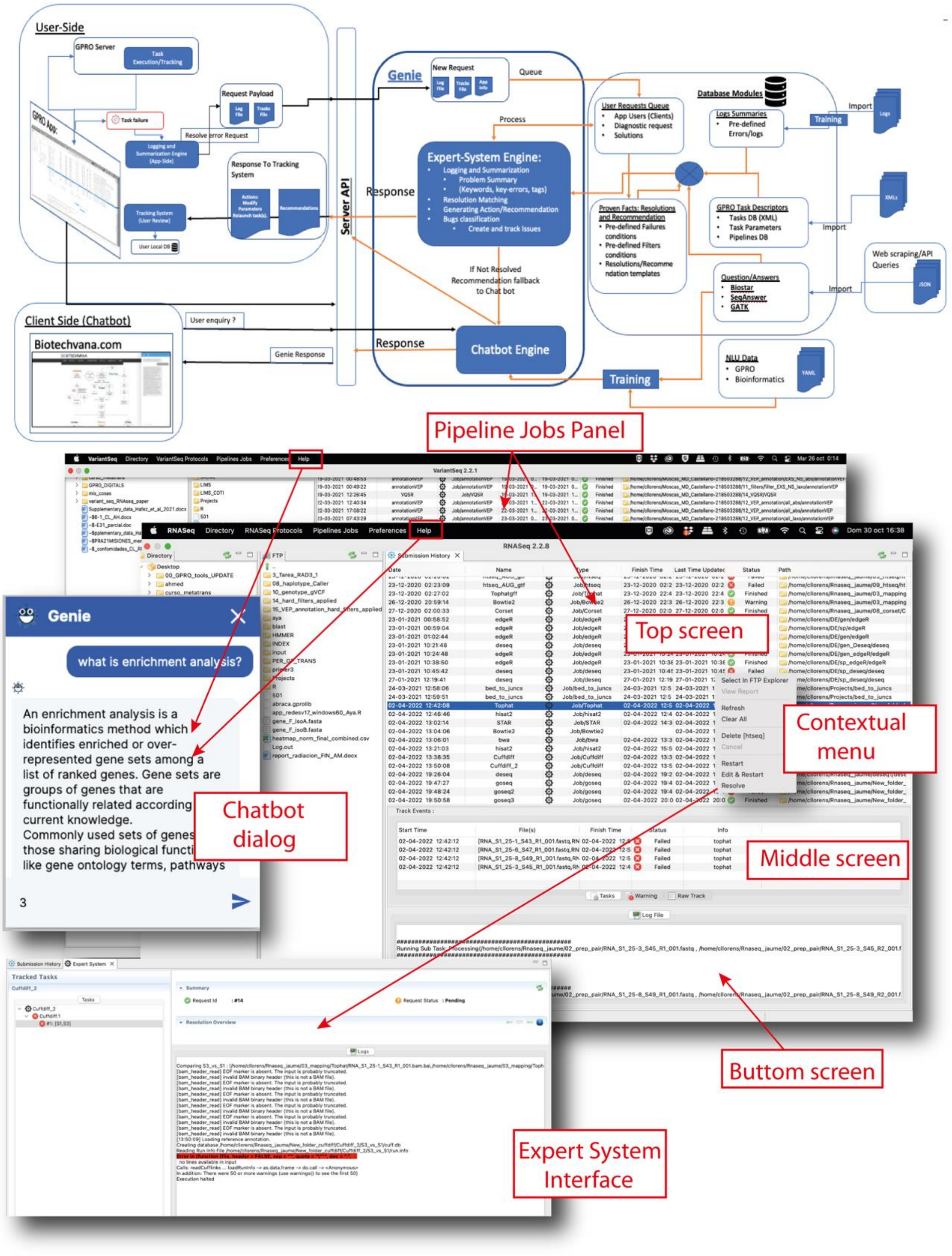

**Figure 5.- user support system.** Schematic of the GENIE AI system. The chatbot and expert system engines, and their knowledge databases are hosted on a remote GPRO server that communicates with the client applications and GSS via API. Each application presents a dialog available in the Help section of the Top Menu where the chatbot can be asked questions. Each application also has a pipeline jobs panel, which is a dynamic interface that summarizes all jobs summitted to the GSS and that provides information across three screens (Top, Middle, and Bottom) about the status of each specific job. Specifically, if the job finished correctly (green icons), had some warnings (orange icons), or failed (red

icons). By right clicking on the panel, a contextual menu will appear to provide tasks to manage the panel (described in text) and the expert system (show at the bottom of the figure).

With the contextual menu, the user can manage options regarding a specific job. For example, in case of a failed job, the user can re-run the analyses using the option *“Edit & Restart"* editing first the settings and parameters of the analysis. If the issue persists the user can access the expert system and try to search a recommendation (if available) about how to solve the issue.

## Discussion

We have developed a client side & server side solution within the GPRO project to perform comparative transcriptomics and variant analysis using CLI tools via GUI environments. The client side part of this solution consists of two applications named “*RNASeq”* and *“VariantSeq”*, both with cloud and desktop executables. Each client application provides a customizable protocol with distinct pipeline or workflows according to the topic (RNA-seq or variant-seq), two modes of execution (Step-by-step and Pipeline-like), and an interactive AI system for troubleshooting. The server side is what we call the GSS a bioinformatic server infrastructure that host the CLI tools and other dependencies needed to run the analyses by the client applications. The GSS is distributed as a docker container image and is easy to install on a remote server or PC.

Comparing “*RNASeq”*, *“VariantSeq”,* and the GSS to other platforms for analysis or resequencing of NGS data is not straightforward as different platforms vary considerably in terms of functionality and features. Nevertheless, we will still attempt to provide the reader with an appropriate summary for our solution relative to other comparable tools. With this in mind, we selected the Galaxy Project [5] as an example of academic Front-end & Back-end solution and CLC of QIAGEN (https://digitalinsights.qiagen.com/) as an example of a proprietary platform. The first aspect to highlight about *“RNASeq”, “VariantSeq”,* and the GSS is that they are publicly available resources as the distinct GUI implementations provided by Galaxy, while CLC is distributed under payment licenses. Respect to scope and performance, our solution is closer to Galaxy than to CLC as our major aim (in the context of the GPRO project) is to provide GUI environments to manage CLI software with the same efficiency and accuracy as if they were running directly from the command line. In contrast, CLC is a multitasking platform whose tasks and tools are mainly based on proprietary implementations. Choosing between this kind of implementation or another is therefore dependent on user experience. It is worth clarifying, however, that the state-of-the-art in bioinformatics is usually defined by public CLI software. As such, the implementation of pipelines and/or workflows based on this kind of software is an increasingly widespread practice among bioinformaticians, both academic and industrial. Similarly, most CLC and proprietary platforms also accept external plugins for third-party CLI software. Although Galaxy and GPRO are contextually similar (provide GUI environments to manage CLI software), Galaxy has wider extent of options than GPRO because is a consortium that provides GUI implementations for the vast majority of CLI software for omic data analysis and not only for differential expression and variant analysis. However, GPRO differs from Galaxy in that our solutions reach to the end-users, fully assembled and ready for use. Along these lines our solution is closer to CLC than to Galaxy in terms of framework implementation. As such CLC offers customized and intuitive solutions (toolkits) for genomics and transcriptomics with distinct tasks for comparative transcriptomics and variant analysis. Both GPRO and CLC applications have been created based on Java frameworks with distinct implementations (desktop, web or cloud, workbench, etc) that are more robust in terms of operativity and framework security than the typical Front-end & Back-end solutions like those typically provided by Galaxy as they are web modules that must be assembled to create the final solution. Therefore, while Galaxy is generally suitable when you are or have support from professional bioinformaticians who can manage and adapt several distinct Galaxy modules to achieve a specific solution. Because of this reason, some end-users, especially those interested in framework friendliness and robustness may prefer platforms based on desktop applications built from conventional (Java, C++, Python, etc.) frameworks instead of web implementations. Of course, this is a question of user experience and/or to the objectives of a bioinformatic resource. While some users may prefer desktop applications, some others will prefer web implementations to locally manage their solutions via web implementations. Logically, if a solution is managed online because is hosted in a remote server with multiple users the web implementation makes more sense. However, Java frameworks like those used to implement platforms like GPRO or CLC can be used to implement desktop versions but also cloud versions that can be deployed in web servers. In fact, we provide desktop and cloud executables of "*RNASeq*" and *"VariantSeq"* (for more details, see the section below "*Data Availability Statement*").

In our opinion, *RNASeq", "VariantSeq",* and GSS combine the user-friendliness, robustness, and security of desktop applications with the operability and versatility of front-end & back end solutions. *“RNASeq”* has been tested successfully in multiple transcriptomics studies using distinct reference sequences and several experimental backgrounds [49,51-53]. *VariantSeq”* has also been validated in several studies of SNP/Indel analysis performed under different experimental contexts (see for example [50,54,55]). Two tutorials (one for “*RNASeq”* and the other for *“VariantSeq”*) have been adapted and presented here to provide users with training material to familiarize themselves with each application. Our solution is also accompanied by an AI system that provides interactive support through an experimental AI system in continuous progress. In this article, we have introduced the two *“RNASeq”* and *“VariantSeq”* applications and the GSS, but we are preparing new publications for other applications of the suite (see the Introduction of this article for more details). We are also planning future implementations for *“RNASeq”, “VariantSeq”,* and the GSS. In the case of *“RNASeq",* we aim to implement additional steps and tools to allow new pipelines and workflows to analyze single-cell RNA-seq and dual RNA-seq data. Regarding *“VariantSeq”,* we want to integrate new steps and tools for the analysis of copy number variations (CNV) and other mutations as well as new tools for filtering, prioritization, and annotation of variants. With respect to the GSS, we are preparing a new release of its docker with multiple user capabilities (the current docker version is limited to one or two users). As such, the GSS will require periodic updates to integrate the new releases and its infrastructure will be progressively increasing in complexity and size. Thus, it is likely that we will eventually split GSS into one docker per application to maintain the user-friendliness of this resource.

**Supplementary material**

**Supplementary file S1.- Pipeline connection settings.** To link "*RNASeq"* or "*VariantSeq"* to the GSS, users need to go to "Pipeline connection settings" under the "Preferences" tab in the Top menu and configure the following settings as if the GSS is installed in a remote server: 1) User email address to receive notifications from the GSS if the latter is installed on a remote server; 2) IP address of the server; 3) Port for the SSH; 5) Username and password of the server for accessing the server. If the GSS is installed locally on the user's PC, the only requirement is to check the option "Run GPRP server locally using Docker" at the bottom of the interface. In the user has been connected to the server, a successful connection notice will be shown when clicking on the tab "Test connection settings".

**Supplementary file S2.- GUI implementation for a CLI tool in *"RNASeq".*** The figure shows the GUI for Cufflinks. These interfaces are usually divided into two blocks. One for declaring the input and output data and another for configuring the options and parameters of the CLI tool invoked by the interface.

**Supplementary file S3.- GUI implementation for a CLI tool in*"VariantSeq"*.** The figure shows the interface for the Base Quality Score Re-calibration (BSQR) command of GATK. Interfaces are usually divided in two blocks, one for input/output data and the other for parametrization and options.

**Supplementary file S4.- Pipeline manager of *"RNAseq".*** Dynamic gif illustrating how to configure and use the pipeline mode in *"RNASeq".*

**Supplementary file S5.- Pipeline manager of *"VariantSeq".*** Dynamic gif showing the implementation and usage of the pipeline mode in *"VariantSeq".*

**Funding:**

This work was supported by the Marie Sklodowska-Curie OPATHY project grant agreement 642095, the pre-doctoral research fellowship from MINECO Industrial Doctorates (Grant 659 DI-17-09134); the Grant TSI-100903-2019-11 from the Secretary of State for Digital Advancement from Ministry of Economic Affairs and Digital Transformation, Spain; the Expedient IDI-2021-158274-a from Ministry of Science and Innovation, Spain; and the ThinkInAzul programme supported by MCIN with funding from European Union NextGenerationEU (PRTR-C17.I1) and Generalitat Valenciana (THINKINAZUL/2021/024).

**Acknowledgements:**

We thank Nathan J Robinson for critical reading and corrections.

**Author Contributions:**

Conceptualization, H.A., B.S., E.A.A., S.J.M., P-S.J., A.V., G.T., and C.L.; Methodology, H.A., B.S., E.A.A., and CL; Software Programming, H.A., B.S., E.A.A., F.R., M.G.; Manuals and tutorial resources, C.R., T-F.M.A., R-R. R, R.F.J., N-C. F., C-G. J.A., T-F. L., G-P. A. and C.L.; Project administration, C.R. and C.L.; Writing and manuscript preparation, H.A., B.S., E.A.A and C.L. All authors read and approved this work.

**Institutional Review Board Statement:**

Not applicable

**Informed Consent Statement:**

Not applicable

**Data Availability Statement:**

RCP (Desktop) and RAP (cloud) executables of *"RNASeq"*, and *"VariantSeq"* are available at:

*"RNASeq"* [ https://gpro.biotechvana.com/download/RNAseq ]

*"VariantSeq"* [ https://gpro.biotechvana.com/download/VariantSeq ]

RCP versions are desktop executables for Windows, Linux, and/or macOS that can be installed in any PC while RAP versions are cloud applications that can be managed with the web browser of any PC.

Manuals of *"RNASeq"* and *"VariantSeq"* are available at:

*"RNASeq"* [ https://gpro.biotechvana.com/tool/rnaseq/manual ]

*"VariantSeq"* [ https://gpro.biotechvana.com/tool/variantseq/manual ]

Tutorials for getting familiar with the installation and usage of *"RNASeq"* and *"VariantSeq"* are accessible at:

*"RNASeq"* [ https://gpro.biotechvana.com/software/RNASeq/RNAseq_tutorial.docx ]

*"VariantSeq"* [ https://gpro.biotechvana.com/software/VariantSeq/Variantseq_tutorial.docx ]

An image of the GPRO GSS Docker is available at the following URL: [https://hub.docker.com/r/biotechvana/gpro], its Installation is easy but requires of a third-party software installer called Docker Desktop that is available at the following URL at https://www.docker.com/products/docker-desktop. To proceed with the installation of the GSS open the Docker Desktop software and execute the following command on its terminal:

```
local_path="/path/to/local_home"

GPRO_USER="myUserName"

GPRO_USER_PASS="myUserNamePass"

> docker run -d -p 80:80 -p 20-22:20-22 -p 65500-65515:65500-65515 -v
/path/to/local_home:/home/gpro_user biotechvana/gpro
```

Please note that the words “myUserName" and "myUserNamePass" above refer to the username and password that the user chooses to access the GSS.

A web version of the chatbot of GENIE is available at the following URL

[ https://gpro.biotechvana.com/genie ]

Fastq files used in the tutorials for RNA-seq and Variant-seq analysis provided as supplementary files S6 and S7 were obtained from the SRA archive at the NCBI [4].

**Conflicts of Interest:**

The authors declare no conflict of interest.

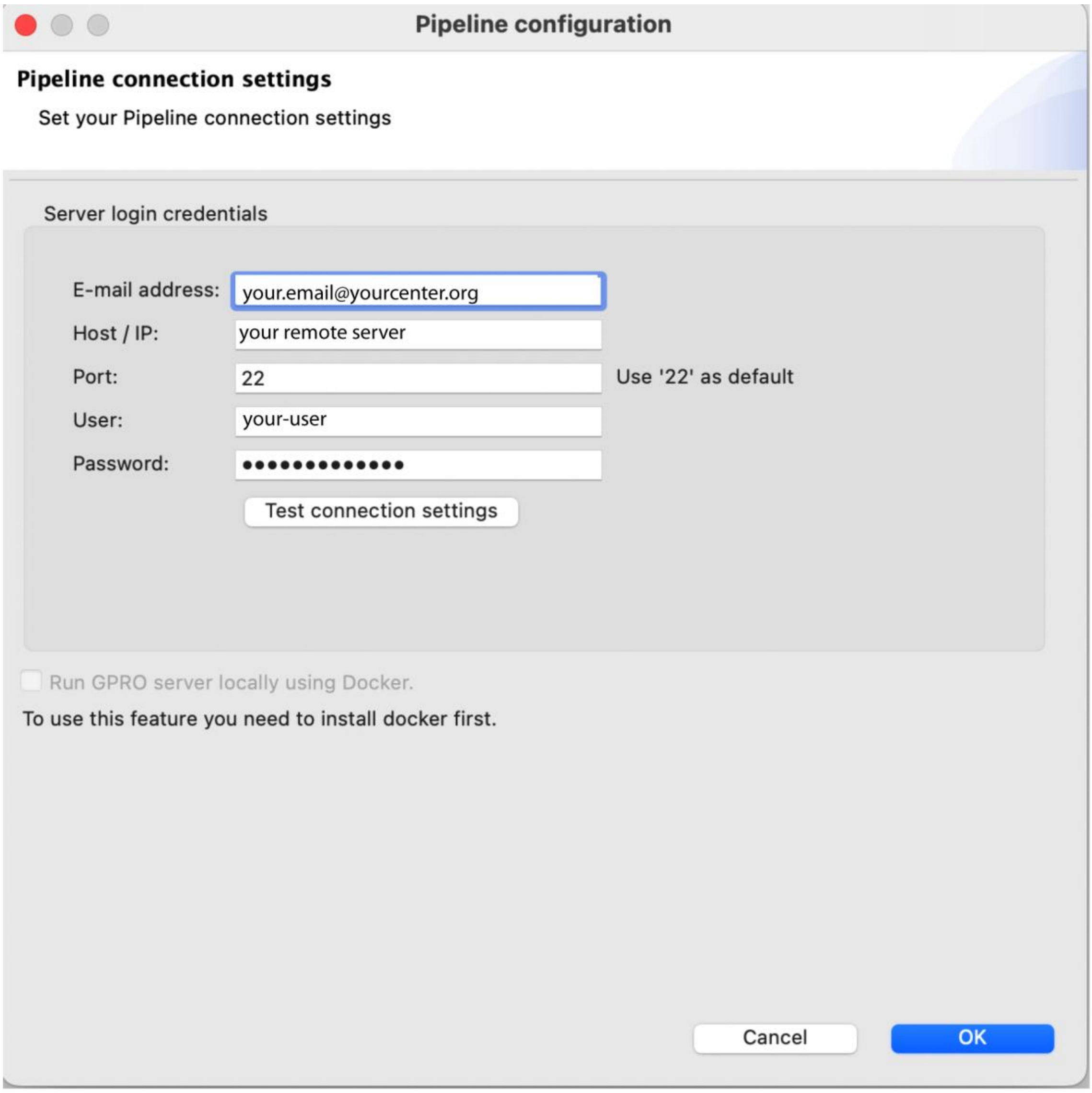
Pipeline configuration
Pipeline connection settings
Set your Pipeline connection settings
Server login credentials
E-mail address:
your.email@yourcenter.org
Host / IP:
your remote server
Port:
22
Use '22' as default
User:
your-user
Password:
Test connection settings
Run GPRO server locally using Docker.
To use this feature you need to install docker first.
Cancel
OK

*Supplementary file S2*

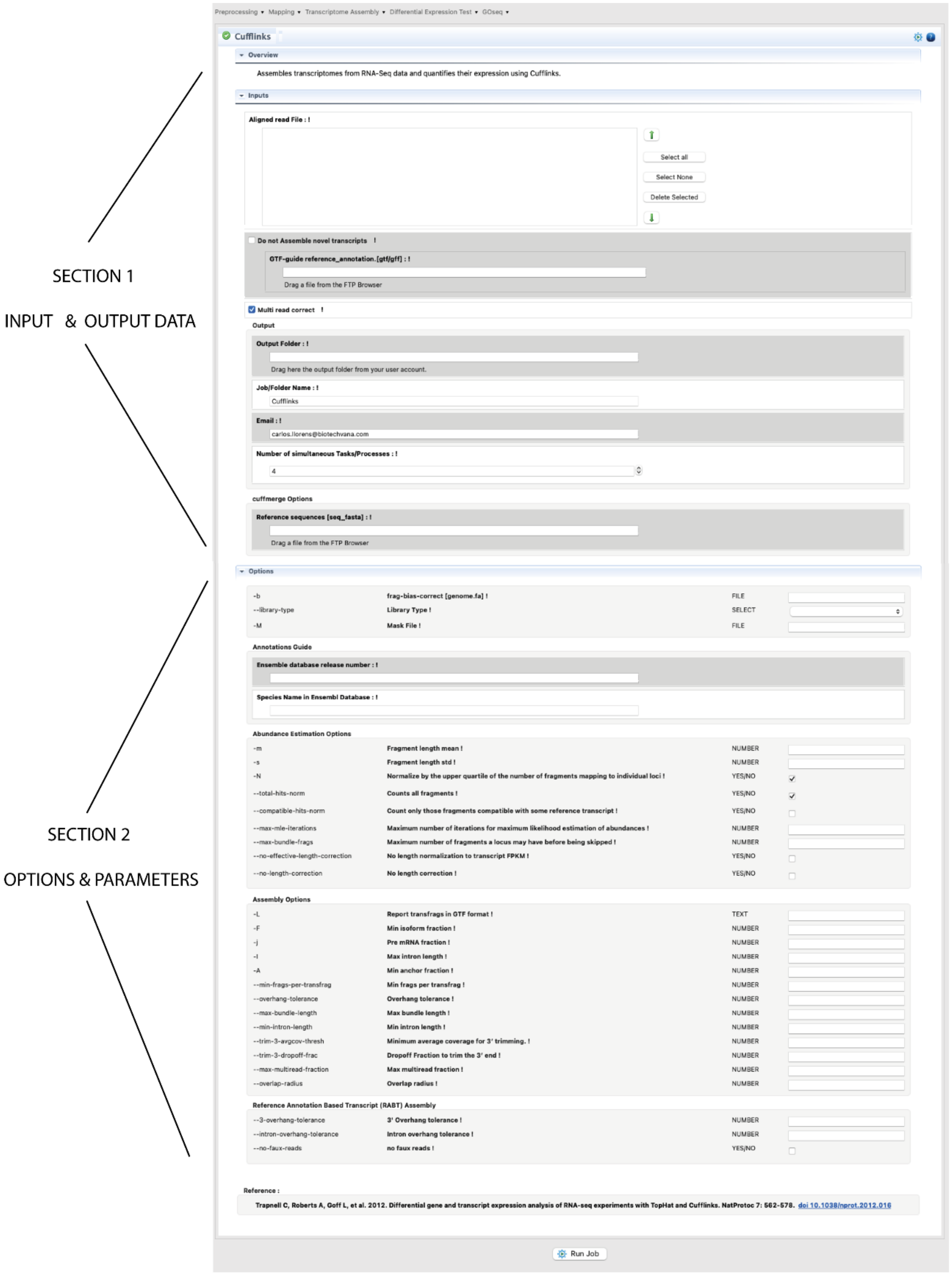

*Supplementary file S3*

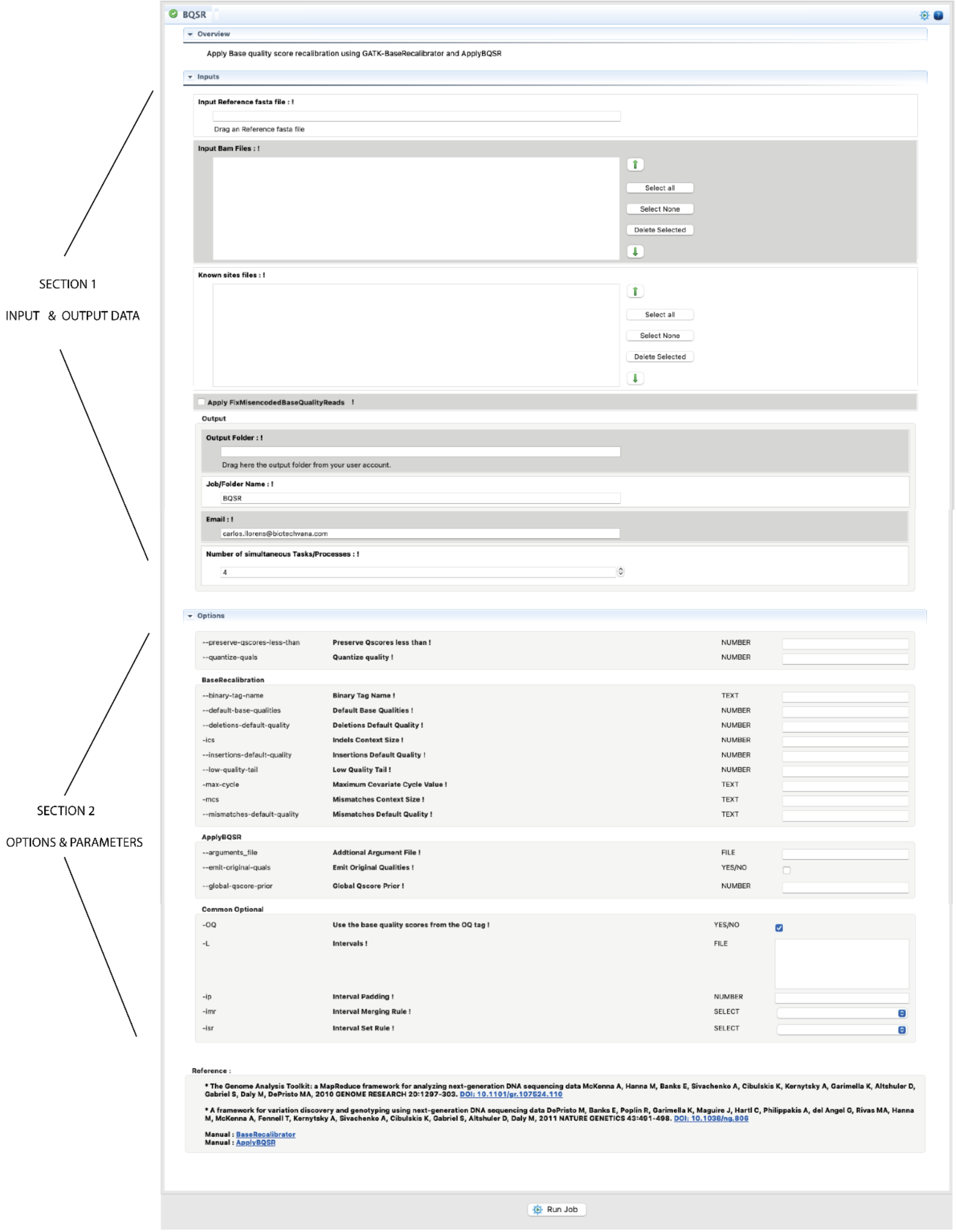

*Supplementary file S4*

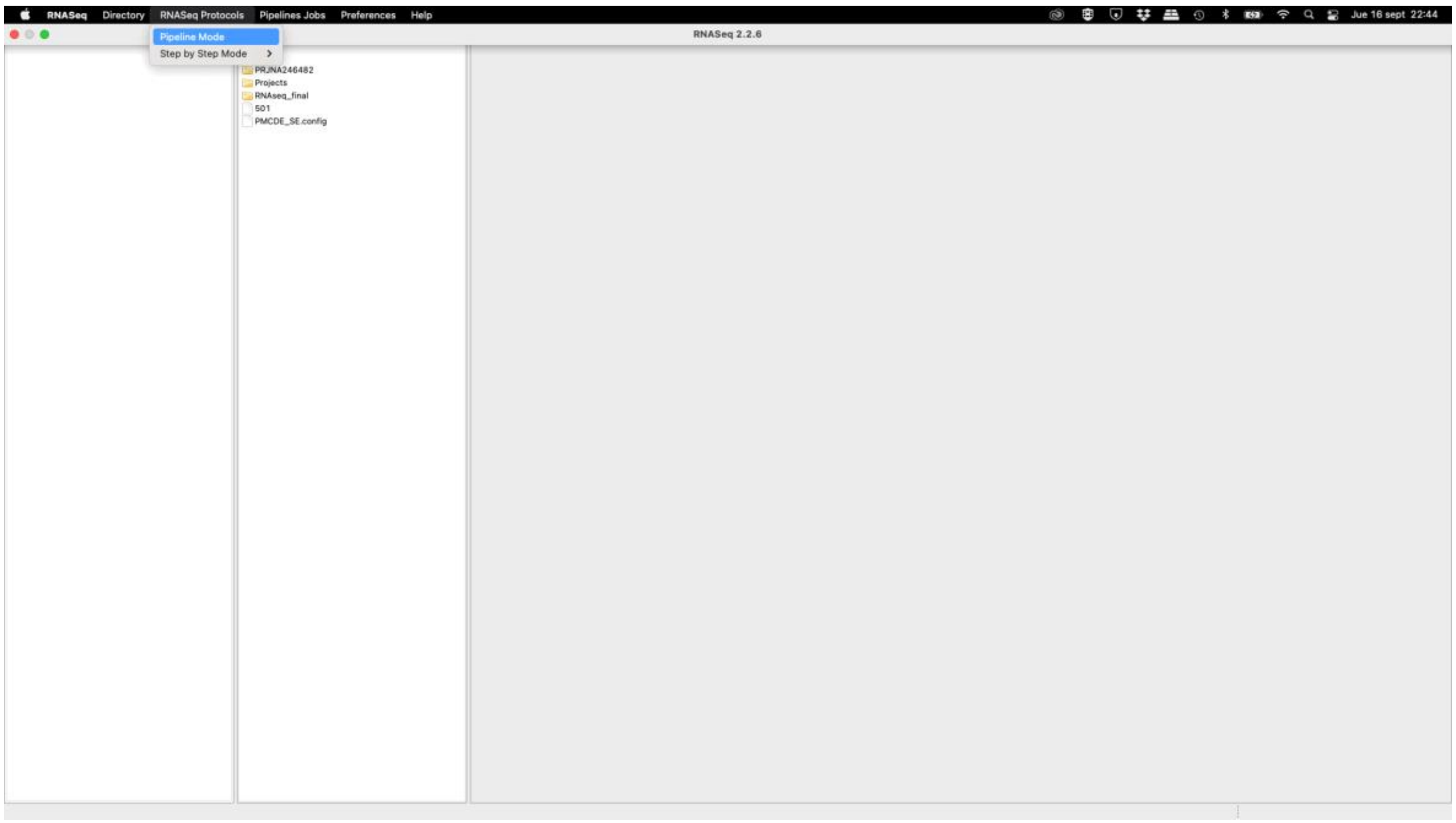

*Supplementary file S5*

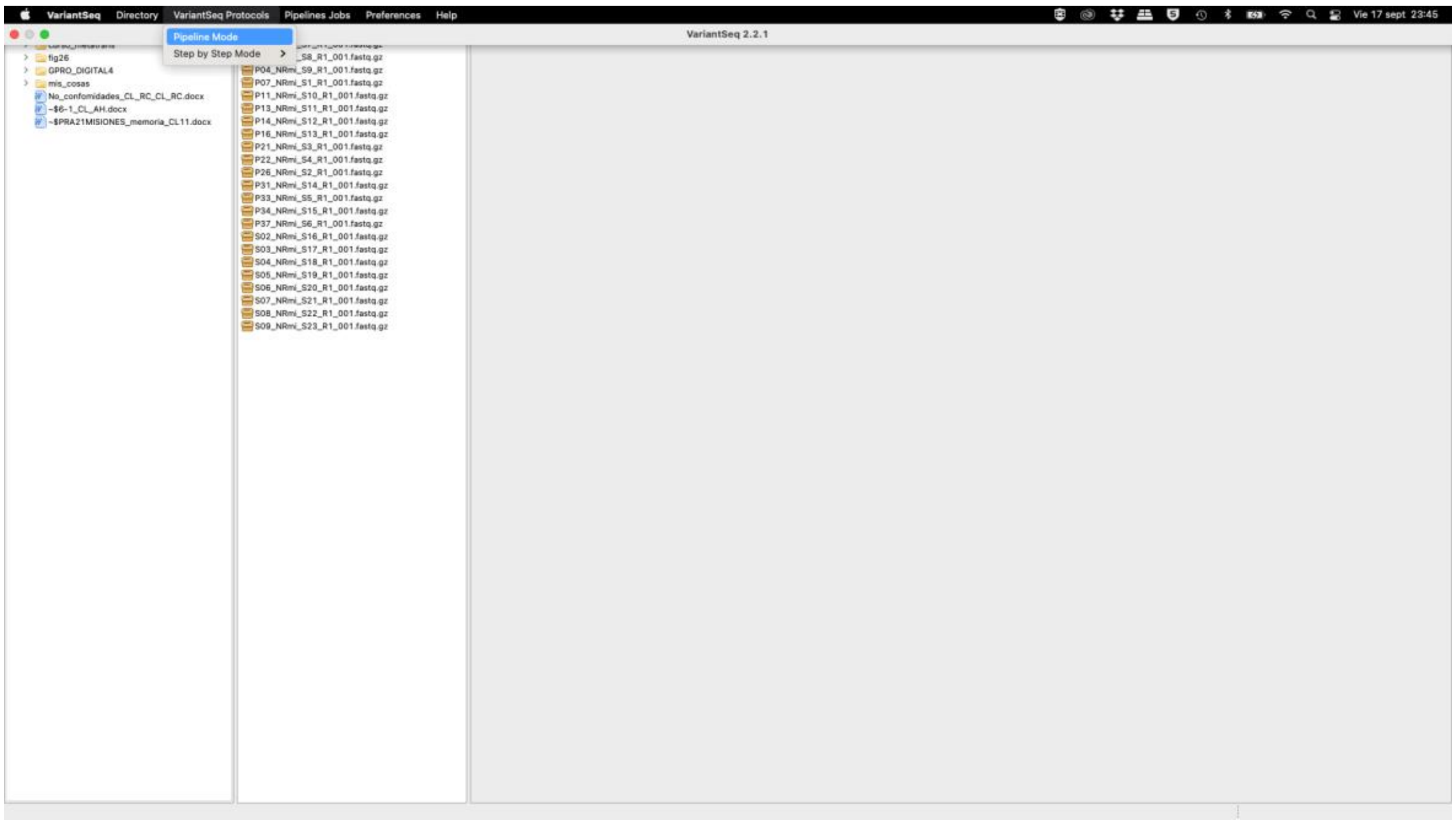